\magnification=1200
\baselineskip=20pt

\def\tile{{\tilde e}}
\def\Ls{\Lambda_s}
\def\me{{\tilde m}_e}
\def\M{{\tilde M}}

\def\gm{\gamma}
\centerline{\bf Probing the SUSY breaking scale at an $e^-e^-$ collider}

\vskip 1 true in

\centerline{\bf Uma Mahanta}

\centerline{\bf Mehta Research Institute}

\centerline{\bf Chhatnag Road, Jhusi}

\centerline{\bf Allahbad-211019}

\vskip .4 true in

\centerline{\bf Abstract}

If supersymmetry is spontaneously broken at a low energy scale then the 
resulting gravitino will be very light. The interaction strength of the
longitudinal components of such a light gravitino to $e-\tile$ pair then
 becomes comparable to that of electroweak interactions. If such a light 
gravitino is present it would significantly modify the cross-section for $e^-_L
e^-_R\rightarrow \tile_L\tile_R$ from its MSSM value. Precision measurement
of this cross-section could therefore be used to probe the low energy
supersymmetry breaking scale $\Lambda_s$.

\vfill\eject

If supersymmetry (SUSY) is spontaneously broken at a low energy scale
 ($\Ls\approx 1-10 $\ \ Tev) then the resulting gravitino 
is expected to be very light ($m_{3\over 2}\approx {F\over M_p}\approx
10^{-4}-10^{-2}$ ev).
The interaction strength of the longitudinal components of such a light
gravitino i.e. the goldstino with fermion-sfermion pair is expected to be
of the order of electroweak couplings [1]. Such a light gravitino could lead
to new and interesting signatures at forthcoming high energy colliders.
It could also significantly modify the collider expectations for many processes
from their corresponding MSSM value [2]. Precision measurement of the 
cross-section for  such processes can therefore be used to set stringent 
bounds on the SUSY breaking scale ($\Ls$).

A high energy $e^-e^-$ collider with the provision for polarising both the
incoming electron beams to a high degree
provides an ideal environment for such studies.
Consider for example the selectron pair production at an $e^-e^-$ collider.
Depending on the polarizations of the incoming electron beams there are three
distinct processes that can be studied namely $e_L e_R\rightarrow
\tile_L\tile_R$, $e_Re_R\rightarrow \tile_R\tile_R$ and $e_Le_L\rightarrow
\tile_L\tile_L$. In the context of MSSM the lowest order contribution to the
first and second processes arise from the t channel exchange of a ${\tilde B}$
[3], whereas the last process receives non-vanishing contributions both from
${\tilde B}$ and ${\tilde W}_3$ exchanges. In this work we shall assume for 
simplicity that the lightest neutralino is gaugino like and more precisely
a bino. The Majorana nature of ${\tilde B}$ and ${\tilde W}_3$ gives rise to
fermion number violating propagtors which is crucial for all the three 
processes to take place. A close examination of the transition amplitudes for
$e_Le_L\rightarrow \tile_L\tile_L$ and $e_Re_R \rightarrow \tile_R\tile_R$
shows that they
arise from the chirality flipping part of the gaugino propagator. They
therefore vanish as the relevant gaugino mass approaches zero. However the 
amplitude for $e_Le_R\rightarrow {\tile}_L{\tile}_R$ arises from the chirality
conserving part of the gaugino propagator and therefore it remains finite in 
the same limit. The longitudinal components of the gravitino behaves as a
Majorana fermion and it also interacts with $e-\tile$ pair through Yukawa
interactions just like the electroweak gauginos. The only difference is that
tha gauge couplings are replaced by the soft gravitino coupling
$e_g\approx { \me^2\over F}$ [4]. Here ${\sqrt F}\approx \Ls$ is the
 scale for dynamical supersymmetry breaking.
 Hence it follows from the above discussion
that the existence of a sufficiently 
light gravitino would modify the cross-section for
$e_Le_R\rightarrow \tile_L\tile_R$ but keep the cross-sections for the 
processes $e_Le_L\rightarrow \tile_L\tile_L$ and $e_Re_R\rightarrow \tile_R
\tile_R$ almost unaffected. Precision measurement of the cross-section for
$e_Le_R\rightarrow \tile_L\tile_R$ could therefore be used to set a lower
 bound on the SUSY breaking scale $\Ls$. This however requires that the
selectron mass $\me$ and the bino mass ($\M$) be known with sufficiently 
good accuracy from other studies. The selectron mass and the bino mass can 
be determined from the energy distribution of the electron arising from 
selectron decay.

The Yukawa interactions of bino ($\tilde {B}$) and gravitino ($\tilde {G}$)
with $e-\tile$ pair are given by [4]

$$L_1=[{g^{\prime}\over \sqrt {2}}\bar{\tilde B}P_L e\tile_L^*-
\sqrt {2} g^{\prime}\bar {\tilde B} P_R e\tile_R^* ]+h.c.\eqno(1)$$.

$$L_2=e_g\sqrt {2}[\bar{\tilde G}P_R e\tile_R^* +\bar{\tilde G}P_L e
\tile_L^*] +h.c.\eqno(2)$$.

In the context of MSSM ($e_g\approx 0$) the transition amplitude for
$e_Le_R\rightarrow \tile_L\tile_R$ arises from the t channel exchange
of a bino and is given by $M=M_a+M_b$ where

$$M_a=-g^{\prime 2}\bar {v}(p_2,s_2)P_R{p_1.\gm-k_1.\gm\over t-\M^2} P_L
u(p_1,s_1).\eqno(3)$$

$$M_b=-g^{\prime 2}\bar {v}(p_2,s_2)P_L{p_1.\gm-k_2.\gm\over u-\M^2}P_R
u(p_1,s_1).\eqno(4)$$

Clearly $M_a$ and $M_b$ are the transition amplitudes associated with the
direct and crossed diagrams. Here ($p_1,p_2$) are the momenta of the incoming 
electrons and ($k_1,k_2$) are the momenta of the outgoing selectrons.
Squaring the transition amplitude and summing over the incoming electron
spins we get

$$\eqalignno{\sum_{s_1,s_2}\vert M \vert ^2
&={2g^{\prime 4}\over (t-\M^2)^2}[2p_1.(p_1-k_1)
p_2.(p_1-k_1)-p_1.p_2(p_1-k_1)^2]\cr
&+{2g^{\prime 4}\over (u-\M^2)^2}[2p_1.(p_1-k_2)p_2.(p_1-k_2)-p_1.p_2
(p_1-k_2)^2].&(5)\cr}$$

Note that at very high energy where the electron mass can be neglected there 
is no interfernce between $M_a$ and $M_b$ i.e. the t and u channel amplitudes.
For simplicity in this work we shall ignore any mixing between $\tile_L$
and $\tile_R$ and assume that $\tilde {m}_{e_L}=\tilde {m}_{e_R}=\me $.
The above expression for $\sum_{s_1,s_2} \vert M \vert ^2 $ then becomes after
some algebra

$$\sum_{s_1,s_2}\vert M \vert ^2 =2g^{\prime 4} (ut-\me ^4)[{1\over (u-\M ^2)
^2}+{1\over (t-\M ^2)^2}].\eqno(6)$$

The contribution of a light gravitino exchange to the transition amplitude 
for $e_L e_R \rightarrow \tile_L\tile_R$ is given by

$$\eqalignno{\delta M &=2e_g^2[{1\over t}\bar{v}(p_2,s_2)
(p_1-k_1).\gm P_L u(p_1,s_1)\cr
&+{1\over u}\bar {v}(p_2,s_2)(p_1-k_2).\gm P_R u(p_1,s_1)].&(7)\cr}$$

Assuming that $\delta M$ is small compared to M we can neglect $\vert \delta
M \vert ^2$ compared to $\vert M\vert ^2$. We then obtain 

$$\eqalignno{\sum_{s_1,s_2}\vert M+\delta M\vert^2 &\approx
2g^{\prime 4}(ut-\me ^4)[{1\over (t-\M^2)^2}+{1\over (u-M^2)^2}]\cr
&-8g^{\prime 2}e_g^2(ut-\me^4)[{1\over t(t-\M^2)}+{1\over u(u-\M^2)^2}].&(8)
\cr}$$

Integrating over all directions the total cross-section for 
$e_Le_R\rightarrow \tile_L\tile_R$ to lowest order in $e_g^2$ becomes
$\sigma_{LR}=(\sigma_{LR})_{mssm}+\delta\sigma_{LR}$ where

$$(\sigma_{LR})_{mssm}={1\over 2\pi s}{\sqrt{s-4\me^2}\over \sqrt {s}}
{g^{\prime 4}\over 4}[{a\over b}\ln {a+b\over a-b}-2].\eqno(9)$$

and

$$\eqalignno{
\delta\sigma_{LR}&=-{1\over 2\pi s}{\sqrt {s-4\me^2}\over \sqrt {s}}
{g^{\prime 2}e_g^2\over 2}[{b^2-a^2\over b(c-a)}\ln {a+b\over a-b}\cr
&-{b^2-c^2\over b(c-a)}\ln {c+b\over c-b}-2].&(10)\cr}$$

In the above $a=\me ^2-\M^2-{s\over 2}$, $b={\sqrt{s}\over 2}\sqrt{s-4\me^2}$
 and $c=\me ^2-{s\over 2}$. A sufficiently light gravitino
would therefore lower the cross-section $\sigma_{LR}$ from its MSSM value.
We find that unless the selectron mass is too close to the threshold
the MSSM contribution to the cross-section is quite large. For example
$\me =150 $ Gev and $\M =100 $ Gev yields a cross-section of 960 fb.
With an integrated luminosity of 50 fb$^{-1}$ per year we therefore expect 
around 48000 events. The statistical error in the cross-section would therefore
be about .4\% which can be further be reduced by increasing the luminosity or 
the running time.
The gravitino contribution to $\sigma_{LR}$ is bounded by the difference
between the experimental value and the MSSM contribution. We therefore need
to estimate the theoretical systematic error 
in the  MSSM cross-section arising from the uncertainties
in $\M$ and $\me$. The values of $\M$ and $\me$ are constrained by the electron
energy distribution to lie in a narrow  elliptical region with positive 
correlation [5]. The MSSM cross-section decreases with increasing $\M$
or $\me$. The contours of constant $\sigma_{LR}$
 are therefore perpendicular
to the uncertainty ellipse.
If we assume that $M$ and $\me$ are both known with an accuracy of 1\% then 
by using  Eqn.(9) it can be shown that the systematic error in the
MSSM cross-section is about
1.5\% for the central values $\M =100$ Gev and $\me =150$ Gev. Both the
systematic error and the statistical error however decreases with decreasing
superpartner mass.
 The condition ${\vert\delta\sigma_{LR}\vert
\over (\sigma_{LR})_{mssm}}\le .01$ could therefore be used to derive an
 approximate lower 
bound on $\sqrt {F}$ provided $\me$ and $\M$ are around 100 Gev.
We find that for $\M =100 $ Gev and $\me =150$ Gev the SUSY breaking scale 
must be greater than 1.4 Tev so that the gravitino contribution is below the
precision limit for measuring $\sigma_{LR}$. The bound corresponds to
a center of mass energy of 500 Gev.
 A technically better estimate
of the bound can be obtained by using the relation ${S\over \sqrt {S+B}}\le 2$.
Here B is the MSSM background and S is the gravitino contribution to the 
signal. For $\me =150 $ Gev, $\M =100 $ Gev and an integrated luminosity of
50 fb$^{-1}$ we then get $\sqrt {F}\ge 1.24 $ Tev which is close to the bound
obtained by using the relation ${\vert \delta \sigma \vert \over \sigma }\le
.01 $.
 The angular distribution of
the gravitino contribution clearly differs from from that of the MSSM
contribution since the former involves a massless propagator and the
latter a massive one.
 Hence by considering an angular range where the gravitino
contribution is enhanced relative to the MSSM contribution it might be
possible to push the lower bound on $\Ls$ to higher values.
What are the implications of our result on known models of low energy 
dynamical supersymmetry breaking (DSB)?
 Recently a lot of interest has been devoted to
the  construction of models with gauge mediated SUSY breaking (GMSB) [6]
 which 
constitutes an example of low energy DSB. However for the simplest models of 
GMSB the SUSY breaking scale $\Ls$ lies between $10^2-10^5$ Tev. Hence the 
process considered in this work will not be able to probe the SUSY breaking 
scale associated with the simplest versions of GMSB. However there could well 
be other scenarios of DSB with a SUSY breaking scale close to 1 Tev
 and super partners in few hundred Gev range which could fall within the
 sensitivity reach of the process considered in this work. The 
supersymmetry breaking in such models has to be communicated to
the visible sector by some
interactions other than SM gauge interactions.

It should be noted that the selectron pair production cross-section at an
$e^+e^-$ collider can also be used to set bounds on $\sqrt {F}$. At an 
$e^+e^-$ collider depending on the polarization of the incoming electron
beam there are two distinct cross-sections that can measured 
namely $\sigma_L$ and $\sigma_R$. To eliminate the contribution of t channel
$\tilde {W}_3$  the incoming electron beam can be chosen to be RH.
The transition amplitude for $\sigma_R$ receives contribution from
t channel $\tilde {B}$ exchange and s channel $\gamma$ and Z exchanges.
The contribution of t channel $\tilde {B}$ exchange  has a chirality
conserving and a chirality flipping piece. Clearly the existence of a very
light gravitino modifies only the chirality conserving piece. However
 there are several advantages in using the $e^-e^-$ collision mode
for probing the SUSY breaking scale instead of the $e^+e^-$ mode.
Firstly in the context of MSSM at an $e^-e^-$ collider,
 $\sigma_{LR}$ gets contribution
only from t channel $\tilde {B}$ exchange. However at an $e^+e^-$
collider $\sigma_{R}$ gets contribution from t channel $\tilde {B}$ exchange
as well as s channel $\gamma $ and Z exchanges. 
 The
analytical expression 
for $\sigma_{R}$ at an $e^+e^-$ collider is therefore
much more complicated than that of $\sigma_{LR}$ at an $e^-e^-$ collider.
Secondly the backgrounds to selectron pair production at $e^-e^-$
collider are very small. Most of the major backgrounds to selectron pair
 production present in $e^+e^-$ are absent in $e^-e^-$ mode. For instance
the W pair production and chargino pair production are prohibited
by fermion number conservation. The remaining $e^-\nu W^-$ background
originating from the LH incoming electron beam
can be suppressed by imposing suitable kinematic cuts on the
transverse energy of the final state electron. These remaining
backgrounds can be calculated and subtracted from the total cross-section 
so as to reduce the total uncertainty in the MSSM contribution.

\centerline{\bf References}

\item{1.} P. Fayet, Phys. Lett. B 69, 489 (1977); 70, 461 (1977);
84, 421 (1979); 86, 272 (1979).

\item{2.} S.Dimopoulos, S. Thomas and J. D. Wells, Nucl. Phys. B 488,
39 (1997); J. A. Bagger, K. T. Matchev, D. M. Pierce and R.-J. Zhang,
Phys. Rev. D 55, 3188 (1997); 
 J. L. Lopez, D. V. Nanopoulos and A. Zichichi, Phys. Rev. Lett.
77, 5168 (1996); J. Ellis, J. L. Lopez and D. V. Nanopoulos, Phys. Lett. B 
394, 354 (1997).

\item{3.} H. E. Haber and G. L. Kane, Phys. Rept. 117, 76 (1985);
H. -C. Cheng, Fermilab-Conf-98/005-T, hep-ph/9801234.

\item{4.} P. Fayet, Phys. Lett. B, 84, 416 (1979); 86, 272 (1979).

\item{5.} H. -C. Cheng, J. L. Feng and N. Polonsky, Phys. Rev. D 56, 6875
(1997); H. -C. Cheng, J. L. Feng and N. Polonsky, Phys. Rev. D 57,, 152 (1998);
 J. L. Feng, Nucl. Phys. B (Proc. Suppl.) 62, 36 (1998).

\item{6.} M. Dine and A. Nelson, Phys. Rev. D 48, 1277 (1993); M. Dine,
A. Nelson and Y. Shirman, Phys. Rev. D 51, 1362 (1995); M. Dine, Y. Nir
and Y. Shirman Phys. Rev. D 53, 2658 (1996).

\end